\documentclass[sigconf]{acmart}

\AtBeginDocument{%
  \providecommand\BibTeX{{%
    \normalfont B\kern-0.5em{\scshape i\kern-0.25em b}\kern-0.8em\TeX}}}

\setcopyright{acmcopyright}
\copyrightyear{2023}
\acmYear{2023}
\acmDOI{XXXXXXX.XXXXXXX}

\acmConference[AIMLSystems 2023]{ The Third International Conference on Artificial Intelligence and Machine Learning Systems}{October 25--28, 2023}{Bangalore, India}
\acmPrice{15.00}
\acmISBN{978-1-4503-XXXX-X/18/06}

\usepackage{listings}
\usepackage{amsmath,amsthm}
\usepackage{tkz-graph}

\begin{document}

\title{Accelerating Causal Algorithms for Industrial-scale Data: A Distributed Computing Approach with Ray Framework}


\author{Vishal Verma }
\email{vishal.verma@dream11.com}
\affiliation{%
  \institution{Dream11}
  \city{Mumbai}
  \country{India}
}

\author{Vinod Reddy}
\email{vinod.reddy@dream11.com}
\affiliation{%
  \institution{Dream11}
  \city{Mumbai}
  \country{India}
}

\author{Jaiprakash Ravi}
\email{jaiprakash.r@dream11.com}
\affiliation{%
  \institution{Dream11}
  \city{Mumbai}
  \country{India}
}


\begin{abstract}
  The increasing need for causal analysis in large-scale industrial datasets necessitates the development of efficient and scalable causal algorithms for real-world applications. This paper addresses the challenge of scaling causal algorithms in the context of conducting causal analysis on extensive datasets commonly encountered in industrial settings. Our proposed solution involves enhancing the scalability of causal algorithm libraries, such as EconML, by leveraging the parallelism capabilities offered by the distributed computing framework Ray. We explore the potential of parallelizing key iterative steps within causal algorithms to significantly reduce overall runtime, supported by a case study that examines the impact on estimation times and costs. Through this approach, we aim to provide a more effective solution for implementing causal analysis in large-scale industrial applications.
\end{abstract}

\ccsdesc[500]{Computer systems organization~Cloud computing}
\ccsdesc[300]{Computer systems organization~Machine Learning}
\ccsdesc{General and reference ~ Performance}

\keywords{causal  machine learning, distributed computing,  observational study}

\maketitle

\section{Introduction}
The industry has shown increasing interest in the field of causal inference in recent years, as it offers the potential to make data-driven decisions that can significantly impact a business. This is particularly relevant for large companies like Dream11 \cite{D11}, which operate on a complex scale of data, and thus relying solely on AB \cite{kohavi_tang_xu_2020} testing for causal inference is not always feasible. As a result, causal studies have become essential in the data-driven decision-making process, as they help companies shape their future business strategies. However, while traditional AB testing \cite{ab_netflix, churn_ML} is helpful, many user-related questions cannot be answered since it is not possible to intervene with users in many of these, making them even more crucial to the company's success.
Several libraries have been introduced for causal discovery and inference (Kalainathan and Goudet, 2019; Sharma and Kiciman, 2020; Beaumont et al., 2021)\cite{kalainathan2019causal,dowhy-sharma,guo2023causalvis} due to the importance of this issue. Nevertheless, these libraries have certain restrictions and limitations when it comes to large-scale of data having hundreds of covariates and confounders.

In this study, we show how we employ a distributed computing framework to create our own causal inference platform (NEXUS), which helps us overcome these difficulties and accomplish the following objectives:
\begin{itemize}
\item A substantial parallelism
\item Scalable and quick tweaking of the hyperparameters
\item Cost  optimizations
\end{itemize}
The purpose of this study is to illustrate and investigate the extent to which such objectives are feasible. We also present a prototype Orthogonal Machine Learning implementation by scaling the EconML Orthogonal ML algorithm with Ray.

\section{PRELIMINARIES}

\subsection{Observational Causal Inference (OCI) : Setup and Assumptions}
For ease of exposition, we consider a situation where our treatment ($t_i$) is binary. Let $\{x_i,t_i,Y_i \}$ represent the data of user $i$, where $x_i$  is a multidimensional vector of confounders, $t_i \in \{0,1\}$  is a binary treatment and $y_i$ is the corresponding outcome of interest. Let $Y(t_i= t)$ represent the outcome of user $i$ under the treatment $t$. If $t$ is the actual treatment received by $i$ then $Y(t_i= t)$ is the actual outcome, otherwise it is the potential (counterfactual) outcome of user $i$ under the treatment. Of course, the fundamental problem in causal inference is that for each user, we only observe either $Y(t_i= 0)$  or $Y(t_i= 1)\footnote{We work under the Rubin Causal Model for now}.$
For the sake of exposition, we consider two widely used estimands of interest, the Average Treatment Effect (ATE), and Conditional Average Treatment Effect (CATE).
\begin{gather}
   \label{eq : 1} \tau_{t_1,t_0} = ATE(t_1, t_0) = E[Y(t_i = t_1) - Y(t_i = t_0)] \\
   \label{eq : 2} \tau_{t_1,t_0}(x) = CATE(x,t_0, t_1) = E[Y(t_i = t_1) - Y(t_i = t_0)|X=x]
\end{gather}
\textit{Note: $\tau_{t_1,t_0} = E_{X}[\tau_{t_1,t_0}(x) | X=x]$}

Of course, as mentioned before, for any user $i$ we will only be able to observe one of the potential outcomes, and therefore the two equations above are not estimable in their current format[Table (\ref{tab: data setting})]. We need to make certain 'identification' assumptions in order to convert the above expressions into terms that are directly estimable through observable data. We now proceed to outline the assumptions under 'selection on observables (unconfoundedness)'.

\subsection{Selection on Observables} \label{Sec 3.1 - SOO}

\begin{table}[t]
  \caption{Fundamental Problem of Causal Inference}
  \label{tab: data setting}
  \begin{tabular}{ccccc}
    \toprule
    $X$ & $T$ & $Y$  & $\widehat{Y(0)}$  &  $\widehat{Y(1)}$\\
    \midrule
    &    0 &   0 &   \textcolor{blue}{0}&  \\
  &    0 &   0 &  \textcolor{blue}{0}&  \\
  &    1 &   1 &  & \textcolor{blue}{1} \\
  &    1 &   0 &  &  \textcolor{blue}{0}\\
  &    0 &   1 & \textcolor{blue}{1} &  \\
  &    1 &   0 &  & \textcolor{blue}{0} \\
    \bottomrule
  \end{tabular}
\end{table}

Identification assumptions are required to go from causal estimands [eq (\ref{eq : 1}), (\ref{eq : 2})] to statistical estimands[such as eq(\ref{eq : identi})] which can be computed from observed data. 
\begin{gather}
     \label{eq : identi} \widehat{\tau}_{t_1,t_0}(x) = E[Y|T=t_1|X=x] - E[Y|T=t_0|X=x] 
\end{gather}

Selection of Observables refers to consistent estimation of causal estimands by controlling for all confounders by making the following assumptions. \\

\textbf{Assumption 1: Consistency}
\begin{gather}
    \label{eq: assn1}
    Y_i = Y(t_i);~\forall i=1,2,...,n
\end{gather}

\textbf{Assumption 2: SUTVA}  
SUTVA essentially requires that a unit's outcome only depends on what treatment was assigned to him/her and does not depend on either what treatment is assigned to other users (no spill-over) or how the treatment was assigned (assignment mechanism).
\begin{gather}
    \label{eq: assn2}
    Y_i \perp T_j; ~\forall j \neq i
\end{gather}

\textbf{Assumption 3: Overlap}
This assumption essentially requires that there is no unit that has an arbitrarily large (closer to 1) or arbitrarily small (closer to 0) probability of being selected for treatment. 
\begin{gather}
    \label{eq: assn3}
    0 < P(T=t|X) < \infty ;~ \forall t \in \mathscr{T}
\end{gather}

\textbf{Assumption 4 : Unconfoundedness a.k.a Ignorability}
No unobserved confounders.
\begin{gather}
    \label{eq: assn4}
    Y(t) \perp T | X; \forall t \in \mathscr{T}
\end{gather}

\begin{figure}[ht]

     \centering

     \begin{tikzpicture}
      \node[draw, circle, fill=blue!30] (T) at (0,0) {\textbf{T}};
      \node[draw, circle, fill=blue!30] (X) at (2,2) {\textbf{X}};
      \node[draw, circle, fill=blue!30] (Y) at (4,0) {\textbf{Y}};
      \node[draw, circle, fill=blue!30] (U) at (4,2) {\textbf{U}};
    
      \draw[->, thick] (X) -- (T);
      \draw[->, thick] (X) -- (Y);
      \draw[->, thick] (T) -- (Y) node[midway, above] {?};
      
    \end{tikzpicture}
     
     \caption{U are unobserved entities. Assumption 4 means that there is no causal link between U and the observed data.}
     \Description{}
     \label{fig: Sec 2 - SOO Graph}
\end{figure}
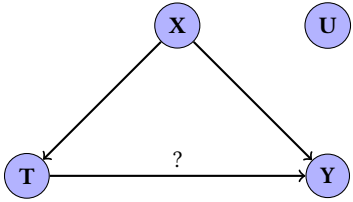

We now show how each of these assumptions is necessary for identification in a setting with no interference [SUTVA holds true]:
\begin{proof}
    \begin{align*}
        \tau_{t_1, t_0} &= E_{X}[E[Y_i(T_i = t_1) - Y_i(T_i = t_0)|X]] \\
    &= E_{X}[E[Y_i(T_i = t_1)|X] - E[Y_i(T_i = t_0)|X]]\\
    &= E_{X}[E[Y_i(T_i = t_1)|T,X] - E[Y_i(T_i = t_0)|T,X]] \text{ [eq(\ref{eq: assn4})]}\\
    &= E_{X}[E[Y|T=t_1,X] - E[Y|T=t_0,X]] \text{ [eq(\ref{eq: assn3})]}
    \end{align*}    
\end{proof}
Note that 
\begin{align*}
       E[Y|T=t,X] &= \int\displaylimits_{y} y f(y|T=t, X)dy \\
    &= \int\displaylimits_{y} y \frac{f(y,T, X)}{P(T=t|X)f(X)}dy\\
\end{align*} 
This necessitates the Overlap assumption so that the conditional expectation is finite.
These assumptions allow methods like 
Covariate Matching\cite{matching_stuart2010}, Metalearners\cite{Kunzel2019}, Doubly Robust(DR) Learners\cite{dr}, Debiased Machine Learning(DML)\cite{dml2018} etc... to estimate the causal effects consistently. One of the major challenges of these methods is that unconfoundedness is fundamentally untestable from data. Although, efforts have been made to assess the sensitivity of the obtained estimates to its \cite{linear_ovb, general_ovb}.

\subsection{Orthogonal/ Debiased Machine learning}
Orthogonal Causal ML (OCML)  is a machine learning method for causal inference that uses orthogonalization to de-bias estimates of causal effects. It works by first regressing all variables on a set of nuisance variables, such as covariates and confounders. The residuals from this regression are then used to estimate the causal effects of interest. OCML has been shown to be effective in a variety of settings, including both observational and experimental data.
The scaling of OCML is more covered in section 5.
Orthogonal statistical learning for treatment effect estimation is a general class of algorithms that work on the following principles:

\begin{itemize}
    \item The $CATE$ function $\theta(X)$ can be thought of as the minimizer of some population risk function $E[l(V,\theta(X),h(V))]$ where $h$ is the nuisance function, and $V=(X,T,Y)$. The loss function is assumed to be Neyman orthogonal. This is achieved by residualising $T$ and $Y$ with respect to $X$ after estimating $h$.
    \item The nuisance function is fit in a cross-fitting manner for each sample to obtain $\hat{h}(V_i); \forall i\in \{1,2,...,n\}$.
    \item If the nuisance function $h$ has been estimated consistently, then the $CATE$ function can be estimated consistently by minimizing the empirical risk function
    $E[l(V,\theta(X),\hat{h}(V))] = \frac{1}{n}\sum\limits_{i=1}^{n}
    l(V_i,\theta(X_i),\hat{h}(V_i))$.
\end{itemize}

\subsection{Distributed computing using Ray}
Ray \cite{moritz2018ray} is an open-source, unified distributed computing framework that makes it easy to scale machine learning (ML) libraries and causal machine learning (CML) tasks. It does this by implementing a unified interface for both task-parallel and actor-based computation
To meet the performance requirements of ML and CML tasks, Ray distributes two components that are typically centralized in existing frameworks, such as Spark \cite{spark}, Dask \cite{dask} etc:
\begin{itemize}
        \item The task scheduler: This component is responsible for scheduling tasks across multiple machines. Ray distributes the task scheduler to ensure that tasks are evenly distributed and that no machine is overloaded.
        \item The metadata store: This component maintains the lineage of computations and a directory for data objects. Ray distributes the metadata store to ensure that data is accessible to all machines.
    \end{itemize}
This distribution of components allows Ray to schedule millions of tasks per second with millisecond-level latency. Ray also provides lineage-based fault tolerance for tasks and actors and replication-based fault tolerance for the metadata store. This means that if a machine fails, Ray can recover the tasks and data that were running on that machine.
Overall, Ray is a powerful and flexible distributed computing framework that is well-suited for scaling ML and CML tasks.

Ray was the superior choice for scaling Orthogonal Machine Learning (OCML) compared to Spark or Joblib due to several key advantages. Ray's lower task overhead and support for distributed state management make it exceptionally well-suited for complex ML tasks that require fine-grained parallelism, as OCML often does. In contrast, Spark and Joblib tend to rely on coarse-grained parallelism, which may not efficiently exploit the full capabilities of modern distributed systems.

Furthermore, Ray's ability to efficiently distribute tasks across multiple machines is a significant advantage over Joblib, which is primarily designed for distributing tasks on a single machine. This allows Ray to harness the full potential of distributed computing resources, making it a more powerful and scalable choice for OCML workloads that require extensive parallelism and distributed processing.
\section{Related Work }

Given the cruciality of the problem, Several implementations have been done tackling the scale and parallelization of CML. Serverless Distributed computing on AWS lambda (Malte S. Kurz et. al)\cite{kurz2021distributed} demonstrate implementation of Double ML using AWS Lambda for parallelization which had shortcomings when it comes to Launch overheads, Limits on memory and run-time, Limits on hyperparameter tuning, Data Transfer, etc. all these limitations can be addressed by our implementation using Ray. CausalAI library (Devansh,et al.)\cite{arpit2023salesforce} also uses Ray Framework for scaling components limited to library implementation. Libraries like EconMl \cite{econml}, CausalML \cite{causalml}, and DoWhy \cite{dowhy-sharma} use libraries like Joblib \cite{joblib} or multiprocessing for multi-threading provide parallelism within a single machine, they do not offer built-in support for distributed computing across multiple machines. Our implementation on the other hand provides efficient implementation and workflow for scaling and serving  Causal ML. 

\section{APPLICATIONS AT Dream11}
As a data-driven company, Dream11  has a culture of experimentation \cite{amit_interview}that enables the exploration of numerous causal inquiries. However, as outlined in Section 1, there are various user-related questions that cannot be answered directly through experimental approaches. Therefore, the importance of observational studies and quasi-experiments becomes evident in addressing these questions.
Dream11's diverse array of categories and the corresponding metrics associated with them present a substantial number of cause-and-effect investigations and connections to be established within the graph.
Furthermore, the utilization of pre-existing algorithms from open-source packages such as CausalML[10] and EconML[6] proves inadequate in handling the voluminous datasets of Dream11 out of the box, which frequently surpasses hundreds of gigabytes in size.
In order to address the aforementioned limitation and successfully scale our causal algorithm, we developed our own in-house causal inference platform called \textit{Nexus}. Leveraging the capabilities of the scaling and distributed computing framework Ray, we extensively evaluated various existing open-source distributed computing frameworks like Apache Spark and Dask \cite{spark,dask}. Ultimately, we determined that Ray was the ideal choice for our specific causal scaling requirements. This comprehensive solution is supported by our internally developed unified analytics and machine learning platform, Darwin [Figure 1].
In terms of capabilities Nexus Offers the following functionalities :
\begin{itemize}
    \item A user-friendly UI for conducting causal analysis.
    \item Functionality to leverage distributed scaling with existing open-source libraries like CausalML, EconML \cite{causalml,econml}
    \item Efficient Deployment and Autoscaling capabilities  using Ray Serve.
    \item Includes integrated validation features such as diagnostic tests, and refutations tests \cite{refutation_amit, schuler2017synthvalidation}.
\end{itemize}

\begin{figure}[h]
  \centering
  \includegraphics[width=\linewidth]{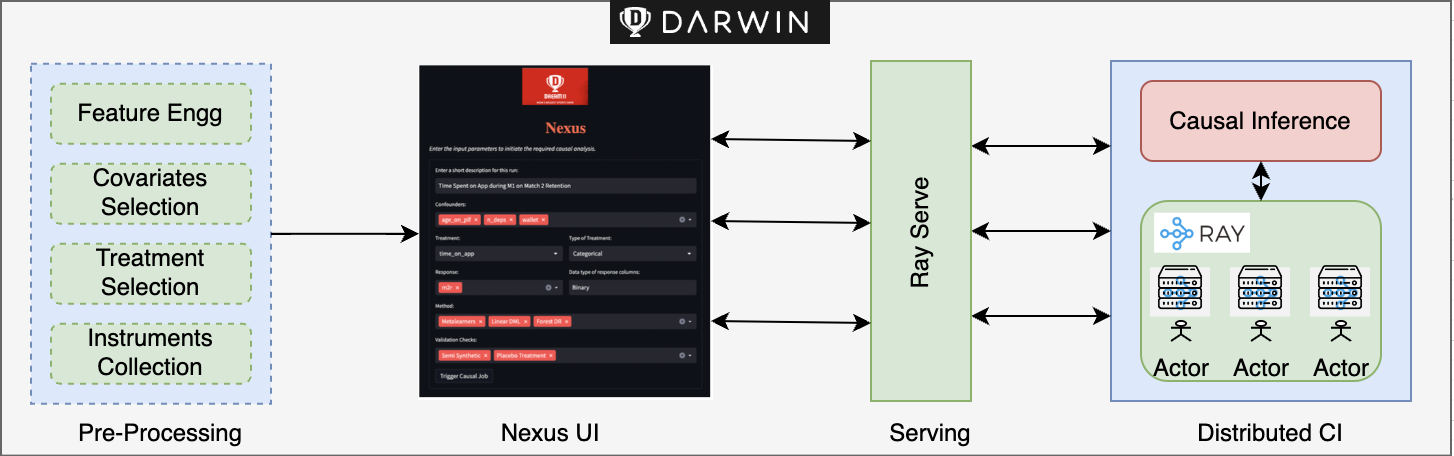}
  \caption{End To End OCI workflow at Dream11}
  
\end{figure}

\section{Case Study: Accelerating OCML}
\subsection{Distributed Crossfitting}
OCML utilizes two ML models (section 2.3). There is a possibility of either one or both models overfitting, which is a common issue with ML models. The solution to this problem involves a technique called cross-prediction or out-of-fold residuals.

Cross-prediction can be performed sequentially or in parallel using multiple threads. However, if we are dealing with large datasets and attempting to scale the approach, the aforementioned methods may not be efficient and may take an exponential amount of time.

To overcome this problem, we can use Ray remote functions, which are called Ray tasks. This allows for remote and asynchronous invocations of each of the K folds simultaneously on separate Python workers.

For demonstration we modified the DML\cite{dml} method from EconML Library to create class \verb|DML_Ray| modified to run cross-fitting in distributed manner as ray tasks having value of parameter cv defined as number of folds

\begin{lstlisting}[language=Python]
import DML_Ray
from econml.sklearn_extensions.linear_model import StatsModelsLinearRegression
from sklearn.ensemble import RandomForestRegressor
from sklearn.ensemble impor import RandomForestClassifier
import ray

import os
runtime_env = {"working_dir": os.getcwd(), "pip": ["dowhy", "econml"]}

#Initialize Ray
ray.init(address='auto', ignore_reinit_error=True, log_to_driver=False,runtime_env=runtime_env)
np.random.seed(123)

#Generating Synthetic Data
X = np.random.normal(size=(1000000, 500))
T = np.random.binomial(1, scipy.special.expit(X[:, 0]))
y = (1 + .5*X[:, 0]) * T + X[:, 0] + np.random.normal(size=(1000000,))

#invoking distributed DML method
est_ray = DML_Ray(
    model_y=RandomForestRegressor(),
    model_t=RandomForestClassifier(),
    model_final=StatsModelsLinearRegression(fit_intercept=False),
    linear_first_stages=False,
    discrete_treatment=True,
    cv=5
)
est_ray.fit(y, T, X=X, W=None)
\end{lstlisting}

\begin{figure}[h]
  \centering
  \includegraphics[width=\linewidth]{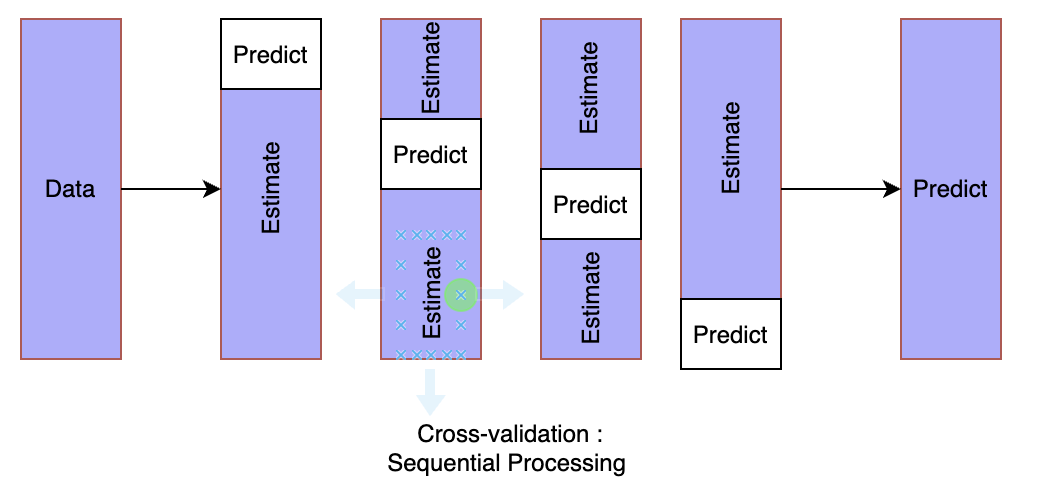}
  \caption{Sequential Cross Validation}
 
\end{figure}

\begin{figure}[h]
  \centering
  \includegraphics[width=\linewidth]{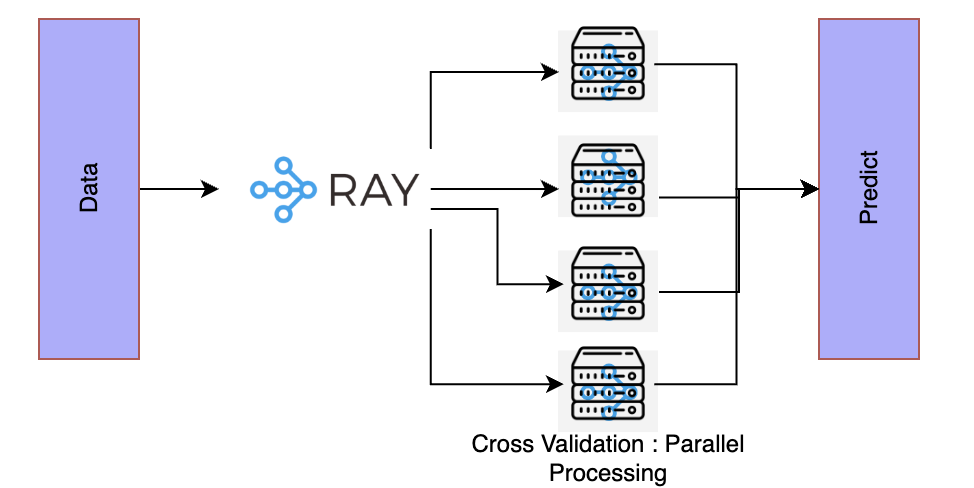}
  \caption{Parallel  Cross Validation using Ray Tasks}
  
\end{figure}

\footnote{ For complete code refer github Gist : https://gist.github.com/vishal-d11/cd886eb6bdff96ad5a04711cb18339ed }
\footnote{ EconML PR : https://github.com/py-why/EconML/pull/800/files }

\subsection{Distributed  Tuning}

In the above discussion, we learned how distributed cross-fitting can be utilized to speed up the Double ML algorithm. However, it is equally important to choose the appropriate \verb|model_y| and \verb|model_t| with the correct hyper-parameters to achieve the best possible results. One solution to this is to leverage the functionality of Ray Tune \cite{raytune} to search through the parameter space and identify the best hyper-parameters. Ray Tune provides a drop-in replacement for the scikit-learn wrapper, which allows for easy integration into the Double ML algorithm.
\verb|est_ray| Above example can be modified to 
\begin{lstlisting}[language=Python]
est_ray = DML_Ray(
    model_y= tune_grid_search_reg(),
    model_t= tune_grid_search_clf(),
    model_final=StatsModelsLinearRegression(fit_intercept=False),
    linear_first_stages=False,
    discrete_treatment=True,
    cv=5
)
\end{lstlisting}

here \verb|tune_grid_search_reg| and \verb|tune_grid_search_clf| is  Ray Tune implementation to search the best estimator and parameter. 

\begin{figure}[h]
  \centering
  \includegraphics[width=\linewidth]{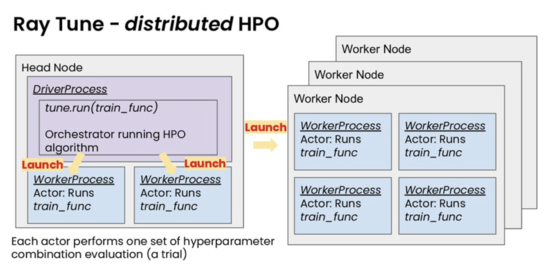}
  \caption{Distributed HyperParam Optimization using Ray Tune (Img source: https://speakerdeck.com/anyscale/fast-and-efficient-hyperparameter-tuning-with-ray-tune?slide=51)}
 
\end{figure}

\subsection{Running Time and Scalability}
We conducted a performance analysis of the EconML implementation of \verb|DML| and our version of \verb|DML_Ray| at varying scales (10k, 100k, and 1Million) of treated units and using approximately 500 covariates generated by a synthetic data generator API sourced from \url{ https://github.com/py-why/dowhy/blob/main/dowhy/datasets.py } \cite{dowhy-sharma}. The results in Figure 5 demonstrate that using Ray to scale the DML algorithm leads to significantly better performance compared to the single-node implementation.

\begin{figure}[h]
  \centering
  \includegraphics[width=\linewidth]{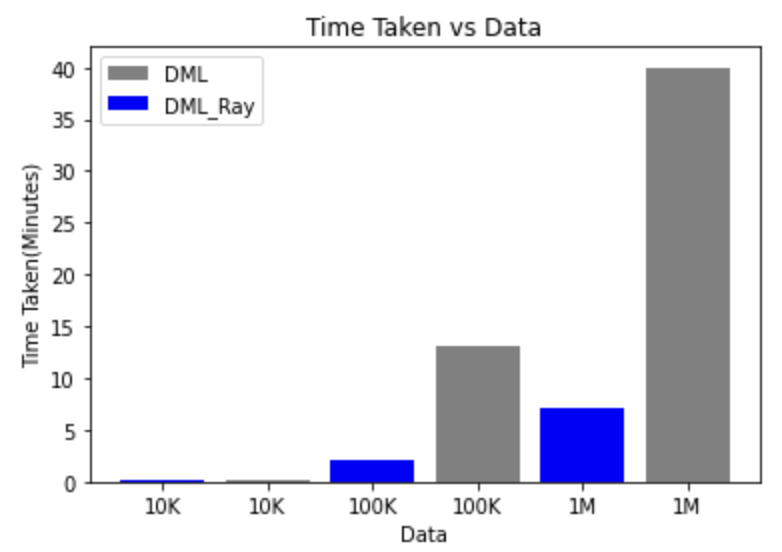}
  \caption{DML vs DMLRay  Runtime on EC2-Highmemory 5 Nodes cluster}
 
\end{figure}

\section{Conclusion and Future Scope}

In conclusion, our paper highlights the importance of developing efficient and scalable causal algorithms to meet the increasing demand for causal analysis in large-scale industrial datasets. We presented a solution for addressing this challenge by leveraging the power of distributed computing through the integration of the Ray framework with causal algorithm libraries like EconML. Our approach involved parallelizing iterative steps in causal algorithms to reduce overall run-time, as supported by the results of our case study. By providing a more effective and scalable solution for implementing causal analysis in industrial applications, we hope our work contributes to advancing the field of causal inference.
In future work, we plan to expand our approach to scale other causal algorithms and look forward to contributing to the open-source community by enhancing the scalability of libraries like EconML and making the approach more accessible to a wider range of users. Moreover, we look forward to scaling up causal discovery algorithms, including those based on Bayesian networks and causal graphical models, using the same principles of distributed computing in the future

\section{Acknowledgements}

We express our gratitude to Hitesh Kapoor and Vinay Jain for setting up the foundation of exploring Ray Framework for data-science causal usecases at Dream11. We would also like to acknowledge the contributions of our Causal inference team members Bihari Pandey, Namita Porwal, Souvik Mohanta, and Nitesh Kumar for their helpful feedback and insights during the course of the development of this paper. We would like to extend our gratitude to Nilesh Patil, Aditya Narisetty Prasad, Rituj Kate, and Darwin team at Dream11  for their assistance in building Nexus. Their support was essential in ensuring the smooth deployment and operation of our platform, and we greatly appreciate their contribution to the success of this project. We acknowledge the use of ChatGPT, an AI language model developed by OpenAI, for assisting in paraphrasing some sections of this research paper.

\bibliographystyle{ACM-Reference-Format}
\bibliography{references}

\end{document}